\documentclass[fleqn,twoside]{article}
\usepackage[headings]{espcrc2}
\readRCS
$Id: espcrc2.tex,v 1.2 2004/02/24 11:22:11 spepping Exp $
\usepackage{graphicx}
\usepackage[figuresright]{rotating}

\newcommand{\AmS}{{\protect\the\textfont2
  A\kern-.1667em\lower.5ex\hbox{M}\kern-.125emS}}

\title{Electroweak and beyond the Standard Model results from HERA}

\author{Z. Zhang\address{Laboratoire de l'Acc\'el\'erateur Lin\'eaire, Universit\'e Paris-Sud 11 et IN2P3/CNRS, BP 34, 91898 Orsay Cedex, France} (for the H1 and ZEUS Collaborations)}
       
\runtitle{Electroweak and beyond the Standard Model results from HERA}
\runauthor{Z. Zhang}

\begin{document}

\begin{abstract}
The latest results from the $e^\pm p$ HERA collider both within the Standard 
(electroweak) Model and beyond are reviewed.
Most of the results are based on the full HERA data sample, 
which corresponds to an integrated luminosity of about $0.5\,{\rm fb}^{-1}$ 
per experiment (H1 or ZEUS).
\vspace{1pc}
\end{abstract}

\maketitle

\section{INTRODUCTION}

The electron-proton collider HERA ended its data taking 
on June 30, 2007. Over its lifetime of more than 15 years of operation, 
it has allowed
each experiment, H1 and ZEUS, to collect an integrated luminosity of
about $100\,{\rm pb}^{-1}$ and $20\,{\rm pb}^{-1}$ in positron-proton
and electron-proton collisions respectively in the first phase (HERA-1)
from 1992 to 2000. In the second phase (HERA-2) from 2003 to 2007, the
positron and electron samples have increased by a factor of $2$ and $10$
respectively. In addition, spin rotators installed at the H1 and ZEUS
interaction regions have provided longitudinal electron and positron
polarization at HERA-2.

The results reviewed in the talk concern both the electroweak (EW) 
measurements in the Standard Model (SM) and searches beyond the SM.
They are based either on the dominant inclusive Deep Inelastic Scattering 
(DIS) processes of neutral and charged current interactions or on rare 
exclusive processes such as single $W$ and lepton pair productions. 
The advantage in the former case is high statistics and in the latter case
clean and well defined final states.

While the emphasis is put on presenting the latest results, 
some early results are
also shown in a few cases in order to illustrate the progress 
that has been made over
the years and the impact of the increased data samples.

\section{DOMINANT INCLUSIVE DIS PROCESSES}

At HERA, both the neutral current (NC) and the charged current (CC) 
interactions can be studied. 
The cross sections of these processes may be written in the following 
simplified form:
\begin{eqnarray}
\sigma_{\rm NC} \hspace{-3mm}&\propto&\hspace{-3mm} \frac{\alpha^2}{Q^4}\Phi_{\rm NC}({\rm PDFs}, v_q, a_q, P_e)\,,\\
\sigma^\pm_{\rm CC} \hspace{-3mm}&\propto&\hspace{-3mm} G^2_F\frac{1\pm P_e}{\left(1+\frac{Q^2}{M^2_W}\right)^2}\Phi_{\rm CC}({\rm PDFs})\,,\label{xscc}
\end{eqnarray}
where $\alpha$ and $G_F$ are the fine structure constant and Fermi constant
for electromagnetic and weak interactions respectively, 
$v_q$ and $a_q$ are vector 
and axial-vector couplings of the $Z$ boson to light quarks $q$ $(u, d)$, 
$P_e$ is the polarization value of the electron beam, $Q^2$ is the
four-momentum transfer squared, $M_W$ the propagator mass of the $W$ boson,
and $\Phi_{\rm NC, CC}$ are the structure function terms for NC and CC
processes respectively, which provide the primary constraint on
the parton distribution functions (PDFs) of the proton.

In NC interactions at high $Q^2$, when $Z$ exchange and $\gamma Z$ 
interference become increasingly important, the NC cross section 
depends both on $v_q$ and $a_q$ and on the beam polarization $P_e$. 
However, these dependencies cannot be factorized, contrary to the $P_e$
dependence of the CC cross section. In addition, from the $Q^2$ dependence 
and the normalization of the CC cross section, one can determine the
propagator mass of the $W$ boson and the Fermi constant $G_F$ respectively.

\subsection{\boldmath The propagator mass of the $W$ boson}

The first evidence of the massive $W$ boson at HERA was obtained from 
the observed deviation of the CC cross section from a linear dependence on the
beam energy based on the very first data 
of the H1 collaboration taken in 1993 corresponding to an integrated
luminosity of $0.35\,{\rm pb}^{-1}$~\cite{h1w93}. 
The electron beam of $26.7\,{\rm GeV}$
in collision with the proton beam of $820\,{\rm GeV}$ resulted in a
center of mass energy $\sqrt{s}$ of $296\,{\rm GeV}$.\footnote{From 1994 on,
the electron beam energy was increased to $27.5\,{\rm GeV}$ and 
from 1998 on, the proton beam energy was also increased to 
$920\,{\rm GeV}$ resulting in $\sqrt{s}=300\,{\rm GeV}$ and 
$\sqrt{s}=318\,{\rm GeV}$.} This is equivalent to 
an electron beam with energy of almost $50\,{\rm TeV}$ hitting a fixed target.

The first quantitative determination of the propagator mass of the $W$
boson, $M_W=76\pm 16_{\rm stat}\pm 13_{\rm syst}\,{\rm GeV}$, 
was derived from the $Q^2$ dependence 
of the CC cross section $d\sigma_{\rm CC}/dQ^2$ with an $e^-p$ data sample of 
$0.54\,{\rm pb}^{-1}$~\cite{zeuswe-95}. Together with the NC cross section
$d\sigma_{\rm NC}/dQ^2$, it was shown that the $Q^2$ dependence of the NC and
CC cross sections are very different. At low $Q^2$ the electromagnetic 
interaction is much stronger than the weak interaction. 
The difference diminishes, however, as $Q^2$ increases, and the cross sections
become comparable when $Q^2$ values reach the $W$ and $Z$ boson mass squared,
demonstrating the EW unification in DIS.

The propagator mass determination in the space-like regime at HERA 
is complementary to other measurements using time-like production of
$W$ bosons at the Tevatron and at LEP and constitutes an important
experimental consistency check of the SM. 

Based on an $e^-p$ ($e^+p$) sample of $0.29\,{\rm pb}^{-1}$ 
($2.7\,{\rm pb}^{-1}$), H1 has performed a similar determination
of the $W$ propagator mass: $M^{e^-p}_W=78^{+11+4}_{-9-3}\,{\rm GeV}$
($M^{e^+p}_W=97^{+18+5}_{-15-10}\,{\rm GeV}$), where the first
error is statistical and the second systematic~\cite{h1w96}. 
It is interesting to note that the precision is better with the $e^-p$ data, 
albeit with a smaller integrated 
luminosity. This is because the $W^-$ ($W^+$) exchange probes mostly the 
$u$ ($d$) quark in the proton resulting in a much larger CC cross section 
in $e^-p$ interactions.

With an increased $e^+p$ data sample of $47.7\,{\rm pb}^{-1}$ taken in
1994-1997 by ZEUS, a simultaneous fit of $M_W$ and $G_F$ was performed 
resulting in $M_W=80.8^{+4.9+5.0+1.4}_{-4.5-4.3-1.3}\,{\rm GeV}$,
where the three errors correspond to the statistical, systematic and PDFs
uncertainty respectively~\cite{zeusw00}. 
When fixing $G_F$ to the world average value~\cite{pdg98}, ZEUS obtained
$M_W=81.4^{+2.7}_{-2.6}\pm 2.0^{+3.3}_{-3.0}\,{\rm GeV}$, where the last
error from the PDFs dominates over the statistical and 
other systematic errors. 

In all above results, the correlation between the EW parameters
with the PDFs was not fully taken into account. A first analysis which 
considered such a correlation was performed by H1, 
using the complete high $Q^2$
NC and CC cross section data and all available low $Q^2$ data from
HERA-1~\cite{h1ew05}. The result on the propagator mass was $M_W=82.87\pm 1.82
^{+0.30}_{-0.16}\,{\rm GeV}$, where the first error is experimental and the
second the model uncertainty~\cite{h1ew05}. 
Within the SM, a much more precise $W$ mass determination, 
$M_W=80.786\pm0.205^{+0.063}_{-0.098}\,{\rm GeV}$, was obtained using the 
SM constraint between $M_W$, $G_F$ and other EW parameters
\begin{equation}
G_F=\frac{\pi\alpha}{\sqrt{2}}\frac{M^2_Z}{(M^2_Z-M^2_W)M^2_W}\frac{1}{1-\Delta r}\,,
\end{equation}
where $M_Z$ is the mass of the $Z$ boson, and $\Delta r$ represents 
radiative corrections which contain among others a quadratic top quark mass 
term and a logarithmic Higgs mass term~\cite{h1ew05}.

The ten-fold increase of the $e^-p$ data from HERA-2 should substantially
improve the final precision of the $W$ mass determination from HERA.

\subsection{\boldmath The light quark couplings to the $Z$ boson}

Using the same HERA-1 data sample, H1 has also performed a combined EW-PDF
fit in which the four light quark couplings $v_u$, $a_u$, $v_d$ and $a_d$
to $Z$ have been determined taking into account their correlation with 
the PDFs~\cite{h1ew05}.
The resulting precision is comparable with those from CDF and the combined
LEP/SLD EW fit. In addition, the HERA data can resolve the sign ambiguity
seen in some of the other determinations.

\begin{figure}[htb]
\hspace{5mm}\includegraphics[scale=0.35]{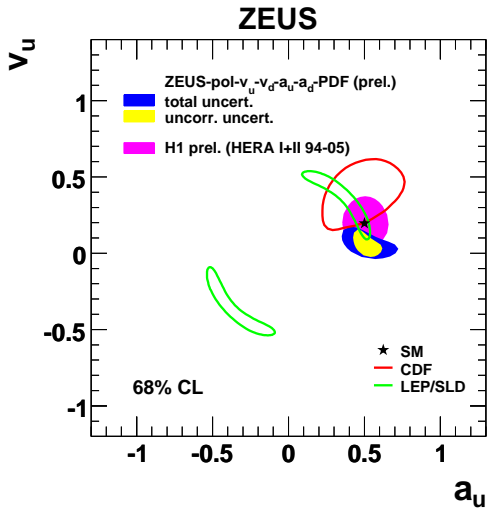}
%\texttt{fig1}
\vspace{-7.5mm}
%\framebox[55mm]{\rule[-21mm]{0mm}{43mm}}
\caption{HERA results at $68\%$ confidence level (CL) on $u$ quark couplings 
to $Z$ in comparison with those from CDF and LEP/SLD.}
\label{fig:ucoup}
\end{figure}

The polarized electron beam at HERA-2 is expected to give additional
sensitivity to these couplings in particular the vector couplings. 
Indeed, including new preliminary NC cross section measurements based on part
of the analyzed HERA-2 data, a new fit
has substantially improved the precision~\cite{hzcoup}. 
The comparison for the $u$ quark couplings with the results of 
CDF~\cite{cdfcoup} and LEP/SLD~\cite{lepcoup} is shown in 
Fig.~\ref{fig:ucoup}.

\subsection{\boldmath Parity violation in NC DIS at high $Q^2$}

\begin{figure}[htb]
\begin{center}
\includegraphics[scale=0.4]{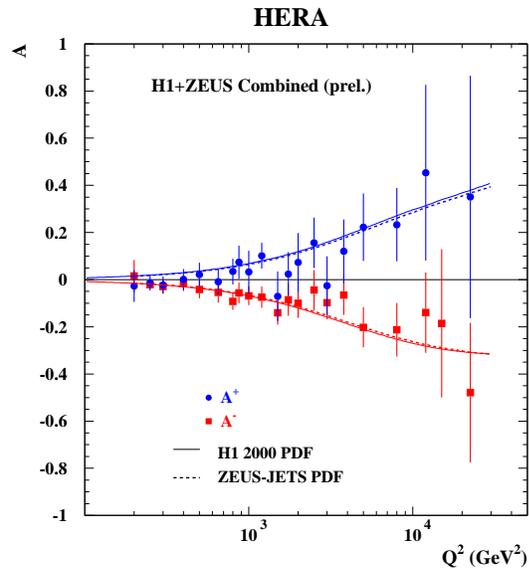}
\end{center}
\vspace{-10mm}
\caption{Measurements of $A^\pm$ by H1 and ZEUS. The error bars
denote the total uncertainty which is dominated by the uncorrelated error 
contributions. The curves describe the theoretical predictions in NLO QCD as
obtained in PDF fits to the H1 inclusive data and to the inclusive and jet 
ZEUS data, respectively. Both fits have been performed using the unpolarized 
HERA-1 data.}
\label{fig:ncasy}
\end{figure}

In the early HERA-1 data with unpolarized electron beam, different
NC cross sections at high $Q^2$ between $e^-p$ and $e^+p$ interactions
were observed, exhibiting the $\gamma Z$ interference and $Z$ exchange
contribution at high $Q^2$. With the polarized electron beam at HERA-2, 
additional differences in the cross sections between the left-handed 
polarization ($P_L$) and 
the right-handed polarization ($P_R$) are expected for the weak interaction
of $Z$ exchange.
A polarization asymmetry $A^\pm$ for $e^\pm$ beams can thus be introduced
\begin{equation}
A^\pm=\frac{2}{P_R-P_L}\frac{\sigma^\pm(P_R)-\sigma^\pm(P_L)}{\sigma^\pm(P_R)+\sigma^\pm(P_L)}\,.
\end{equation}
The preliminary results combining the H1 and ZEUS HERA-2
data are shown in Fig.~\ref{fig:ncasy}~\cite{hzpv}. 
This is the first observation of parity violation in NC DIS at HERA.

\subsection{Limits on finite quark radius and contact interactions}

NC interactions at high $Q^2$ at HERA provide an important test of the SM
at the high energy frontier. Indeed, the SM $Q^2$ spectrum is expected to
be modified if the interacting electron and quark have a finite size:
\begin{equation}
\frac{d\sigma_{\rm NC}}{dQ^2}=\frac{d\sigma^{\rm SM}_{\rm NC}}{dQ^2}
f^2_e(Q^2) f^2_q(Q^2)\,,
\end{equation}
where $f(Q^2)=1-\frac{\langle r^2\rangle}{6}Q^2$ is the electron or quark
form factor, $\langle r^2 \rangle$ being the mean squared radius of 
the EW charge distribution. Assuming a point-like electron, i.e. setting
$f_e\equiv 1$ as it has been well constrained elsewhere~\cite{hsm91}, 
and using the full HERA (preliminary HERA-2 and published HERA-1) data, 
an upper limit on the radius of the light $u$ and $d$ quarks, 
$R_q=\sqrt{\langle r^2_q\rangle}<0.75\cdot 10^{ -18}\,{\rm m}$, 
at $95\%$ CL has been derived by H1~\cite{h1qsize}. 
The limit $R_q<0.62\cdot 10^{-18}\,{\rm m}$ from ZEUS is 
similar~\cite{zeusqsize}.

\begin{figure}[htb]
\begin{center}
\includegraphics[scale=0.37]{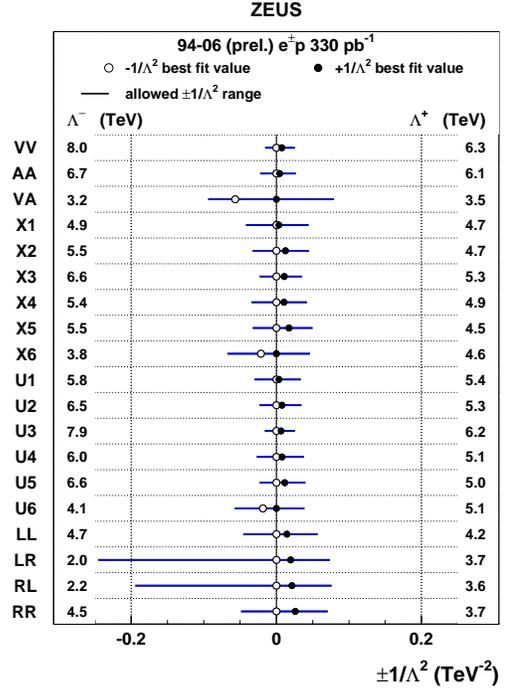}
\end{center}
\vspace{-10mm}
\caption{Confidence intervals of $\pm 1/\Lambda^2$ at $95\%$ CL for
the studied contact interaction scenarios. The numbers at the right
and left margins are the corresponding lower limits on the mass
scale $\Lambda^+$ ($\eta>0$) and $\Lambda^-$ ($\eta<0$) respectively.}
\label{fig:ci}
\end{figure}

New interactions between $e$ and $q$ involving mass scales $\Lambda$ 
above the center of mass energy can also modify the SM NC cross section 
at high $Q^2$ via virtual effects. Such interactions can be modeled as 
four-fermion contact interactions as an effective theory by
additional terms in the SM Lagrangian
\begin{equation}
{\cal L}=\sum \eta^{eq}_{ij}(\bar{e}_i\gamma^\mu e_i)(\bar{q}_j\gamma_\mu q_j)\,,
\label{LCI}
\end{equation}
where the sum runs over electron and quark helicities and quark flavors.
In Eq.(\ref{LCI}), only vector currents are considered as strong limits have 
already been placed on scalar and tensor contact interactions.
The coupling $\eta^{eq}_{ij}$ may be written as 
$\eta^{eq}_{ij}=\epsilon^{eq}_{ij}\frac{4\pi}{\Lambda^2}$, where
the coefficient $\epsilon^{eq}_{ij}$ may take the values $\pm 1$ or zero.
In the HERA contact interaction analyses, 19 different chiral structures
have been considered. The preliminary results from ZEUS are shown in 
Fig.~\ref{fig:ci}~\cite{zeusqsize}. 
The typical excluded scale is $\Lambda>5\,{\rm TeV}$ at
$95\%$ CL. 
Comparable limits were achieved by H1~\cite{h1cthera1}.

\subsection{Total CC cross sections and constraint on right-handed currents}

\begin{figure}[htb]
\begin{center}
\includegraphics[scale=0.37]{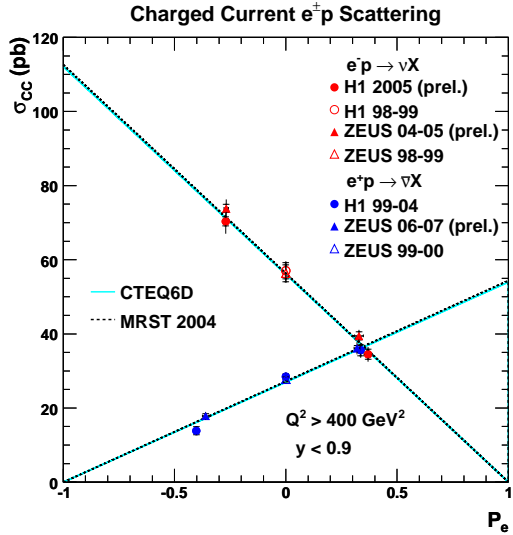}
\end{center}
\vspace{-10mm}
\caption{The dependence of the measured total CC cross sections on 
the $e^\pm$ beam polarization in comparison with the expectations from
CTEQ6D~\cite{cteq6d} and MRST 2004~\cite{mrst}.}
\label{fig:ccxs}
\end{figure}

According to Eq.(\ref{xscc}), the CC cross section of purely weak interactions
is expected to have a linear dependence on the beam polarization $P_e$.
The (preliminary) measurements of the total CC cross section for 
$Q^2>400\,{\rm GeV}^2$ and inelasticity $y<0.9$ with polarized HERA-2 data
are shown in Fig.~\ref{fig:ccxs} together with the corresponding cross sections
of the published HERA-1 data with unpolarized beams~\cite{ccpe}. 
The measurements are in good agreement with the expected linear 
dependence. 

In the SM, only the left-handed $W$ boson participates in the weak 
interaction, and the corresponding cross sections thus vanish for 
$P_e=-1$ ($e^+p$) and $P_e=1$ ($e^-p$). The production of right-handed CC
would result in a deviation from zero at these polarization values. 
The contribution of a right-handed CC can be derived from linear fits 
to the measured cross sections and from extrapolating the results of the
fits to $P_e=\pm 1$. 
The upper limit on the right-handed CC cross sections can be further
translated into lower mass limits on a right-handed $W$~\cite{ccpe}.

\subsection{Other searches in inclusive DIS processes}

Many searches for new particles such as leptoquarks (LQs), squarks 
(the supersymmetric partners of quarks) and 
excited states of fermions have been carried out at HERA, 
since the very beginning of data taking. Both H1 and ZEUS 
have performed their first search using the data of 1992 of 
less than $30\,{\rm nb}^{-1}$~\cite{hz92}.
Indeed, the HERA $ep$ collider is an ideal machine to look for 
resonant production of new particles (LQs, squarks), which couple
via Yukawa couplings directly to a lepton and a parton, up to 
the center of mass energy.~\footnote{Beyond the kinematic limit, the search
sensitivity arises from $u$-channel virtual exchange of new particles coupling
to lepton-quark pairs.} This is in contrast to the pair-production
in $e^+e^-$ at the LEP or in $p\bar{p}$ at the Tevatron collider, 
in which case the mass
reach for a direct search is up to half of the center of mass energy.

The search for the first generation LQs
considers its decay into an electron and a quark or into a 
neutrino and a quark. These LQ decays lead to final states identical
to DIS events. One cannot distinguish them on an event by event basis, but
has to rely on different angular distributions, e.g.\ flat for scalar LQs and
steeply falling for DIS events.

In the history of the HERA running, one of the most exciting moments happened 
in 1997 when both H1 and ZEUS reported an excess of high $Q^2$ events
based on their 1994-1996 data of less than $20\,{\rm pb}^{-1}$~\cite{hz96}.
The excess generated in a short period of time a huge number of 
speculations on possible new physics including resonant production of
LQs or squarks. However, the later high statistics
data samples collected by both experiments did not confirm
the excess, and constraints on LQ production were derived.

\begin{figure}[htb]
\begin{center}
\includegraphics[scale=0.37]{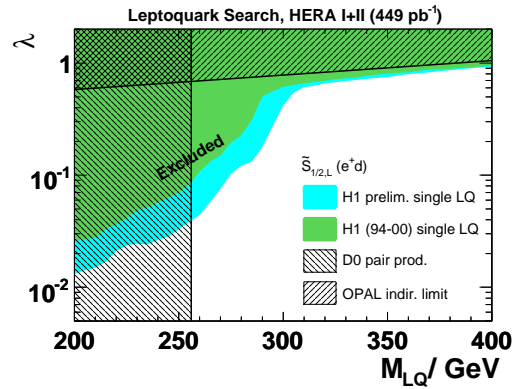}
\end{center}
\vspace{-10mm}
\caption{Example of mass-dependent upper bounds on the Yukawa coupling 
$\lambda$ of a first generation leptoquark coupling to $ed$ 
($\tilde{S}_{1/2,L} (ed)$).}
\label{fig:lq}
\end{figure}

An example for the existing constraints using the full HERA data sample 
from H1 on a scalar LQ, which decays exclusively into an electron and
a quark, is presented in Fig.~\ref{fig:lq}~\cite{h1lq07}. 
The new preliminary limit significantly improves the previous limit from 
HERA-1~\cite{hzlq1}.
For moderate Yukawa coupling values $\lambda$ of about $0.03-0.6$, 
it also extends well beyond the direct limit 
from the Tevatron~\cite{d0lq} and the indirect limit from LEP~\cite{opallq}.

Searches for LQs coupling to first and second generation fermions have also
been performed~\cite{lqlfv}. 
In this case, the final state lepton is different from
the incident electron (lepton flavor violating processes), and the searches
are thus essentially background free. The corresponding constraints are
also more stringent than those on the first generation LQs.

In supersymmetric (SUSY) models where the so-called $R$-parity ($R_p$) is not 
conserved, squarks could be resonantly produced at HERA similar to LQs.
In addition to the LQ-like decays, squarks also undergo decays into
gauginos (the supersymmetric partners of gauge bosons). A dedicated
search requires a large number of final states to be analyzed.
This has been pioneered by H1 in~\cite{h1rpv1}, where the full HERA-1
data have been used to set constraints on SUSY models. 
A similar analysis looking for the SUSY partner of the top
quark has been performed by ZEUS~\cite{zeusstop}.

The observed replication of three fermion families suggests the possibility
that these fermions may be composite particles, consisting of combinations
of more fundamental entities. The observation of excited states of
leptons or quarks would be a clear signal that these particles are composite
rather than elementary. At HERA, excited electrons, quarks and neutrinos
($e^\ast, q^\ast, \nu^\ast$) with masses up to the kinematic limit
of the center of mass energy could be produced directly via $t$-channel 
exchange of a gauge boson.

The full HERA $e^\pm p$ data sample collected by H1 has been used to 
search for excited electrons via the decays $e\ast\rightarrow e\gamma$,
$e^\ast\rightarrow eZ$ and $e^\ast\rightarrow \nu W$~\cite{h1e*}. 
A similar preliminary search has also been performed for excited quarks 
via the decays $q^\ast\rightarrow q\gamma$ and 
$q^\ast\rightarrow qW$~\cite{h1q*}.
The search for excited neutrinos decaying via $\nu^\ast\rightarrow \nu\gamma$,
$\nu^\ast\rightarrow \nu Z$ and $\nu^\ast\rightarrow eW$ has been made
using only $e^-p$ collisions as the corresponding cross section in $e^+p$
is much smaller~\cite{h1nu*}. 
No evidence for excited fermions has been found, and limits on
the characteristic couplings have been derived~\cite{h1e*,h1q*,h1nu*}.

\section{RARE EXCLUSIVE PROCESSES}

The large data sample of HERA has made it possible to use rare exclusive 
processes with well measured and isolated final states for 
performing EW measurements and for searching for deviations
from the SM. Two examples are multi-lepton events with high transverse
momenta ($P_T$) and isolated lepton events with large missing transverse 
momentum ($\not\! P_T$).

\subsection{\boldmath Multi-lepton events at high $P_T$}

Using the full HERA data, H1 has completed~\cite{h1final} a study on 
the multi-lepton events of which an excess was reported earlier based
on HERA-1 data~\cite{h1ml1}. Seven final state topologies
$ee$, $\mu\mu$, $e\mu$, $eee$, $e\mu\mu$, $ee\mu$ and $eeee$ have been 
analyzed.
In all topologies, the predicted event rate agrees with the number of
observed events~\cite{h1final}. However, when a comparison is made 
requiring for the invariant mass of the two highest $P_T$ leptons 
$M_{12}>100\,{\rm GeV}$, an excess is observed in most of the topologies
(Table~\ref{tab:ml}), although the number of observed events remains 
statistically limited. 
Demanding $\sum P_T>100\,{\rm GeV}$ for the leptons,
$5$ $(0)$ events are observed in the $e^+p$ $(e^-p)$ sample with 
$0.96\pm 0.12$ $(0.64\pm 0.09)$ expected. 
Also shown in Table~\ref{tab:ml} are preliminary
results from ZEUS on di-electron and tri-electron samples~\cite{zeusml}.
In both samples no excess has been observed. 
%Therefore the excess is only observed in the $e^+p$ sample.
\begin{table*}[htb]
\begin{center}
\caption{The number of observed events and SM expectations in different 
multi-lepton topologies for $M_{12}>100\,{\rm GeV}$. The numbers shown in 
parentheses correspond to the contribution from the dominant pair production 
in $\gamma\gamma$ interactions.}
\begin{tabular}{|c|c|c|c|c|}
\hline Topology & \multicolumn{2}{|c|}{H1} & \multicolumn{2}{|
c|}{ZEUS} \\ \cline{2-5}
 & Data & SM (pair) & Data & SM (pair) \\
\hline
$ee$ & $3$ & $1.34\pm 0.20$ $(0.83)$ & $2$ & $1.7\pm 0.2$ $(0.9)$ \\ \hline
$e\mu$ & $1$ & $0.59\pm 0.06$ $(0.59)$ & & \\ \hline
$eee$ & $3$ & $0.66\pm 0.09$ $(0.66)$ & $2$ & $1.0\pm 0.1$ $(1.0)$ \\ \hline
$\mu\mu$ & $1$ & $0.17\pm 0.07$ $(0.17)$ & & \\ \hline
$e\mu\mu$ & $2$ & $0.16\pm 0.05$ $(0.16)$ & & \\ \hline
\end{tabular}
\label{tab:ml}
\end{center}
\end{table*}

\subsection{Search for doubly-charged Higgs bosons}

Motivated by the observed excess of the multi-lepton events discussed above, 
a search for doubly-charged Higgs bosons ($H^{\pm\pm}$) has been performed
by H1 using the HERA-1 data~\cite{h1dch}.
The search looks for events with a singly produced $H^{\pm\pm}$ decaying 
into a high mass pair of same charge leptons, one of them being an electron.
No evidence for $H^{\pm\pm}$ production is observed; 
out of the observed multi-lepton events  from the HERA-1 data, 
only one di-electron event is compatible with the 
$H^{\pm\pm}$ signature. Mass dependent upper limits are derived on the 
Yukawa coupling $h_{el}$ of the Higgs boson to an electron-lepton pair. 
Assuming that the doubly-charged Higgs decays only into an electron and
a muon via a coupling of electromagnetic strength 
$h_{e\mu}=\sqrt{4\pi\alpha_{em}}=0.3$, a lower limit of $141\,{\rm GeV}$ 
on the $H^{\pm\pm}$ mass is obtained at $95\%$ CL. 
For a doubly-charged Higgs decaying only into an electron and a tau and 
a coupling $h_{e\tau}=0.3$, masses
below $112\,{\rm GeV}$ are ruled out.

\subsection{\boldmath Isolated lepton events at large missing $P_T$}\label{sec:isol}

Since the observation of the first isolated muon event with large 
$\not\!\! P_T$ by H1 in 1994~\cite{h1isol94}, searches have been 
pursued and extended by H1 and ZEUS to all three types of leptons and 
with increasingly larger data samples~\cite{hziso}.
The results~\cite{h108,zeus08}~\footnote{The preliminary results of H1 shown 
at the workshop have been replaced by the published one. 
The same is true for the results shown in sections~\ref{wprod} and 
\ref{g-search}.} with the full HERA data are summarized in 
Table~\ref{tab:isol}.
\begin{table*}[htb]
\begin{center}
\caption{The number of observed events and SM expectations for
three types of leptons in $e^\pm p$ data samples for $P^X_T>25\,{\rm GeV
}$. The SM signal (dominated by $W$ production) is shown in percentage
in parentheses.}
\begin{tabular}{|c|c|c|c|c|c|}
\hline Dataset & Lepton & \multicolumn{2}{|c|}{H1~\cite{h108}} & \multicolumn{2}{|c|}{ZEUS~\cite{zeus08}} \\ \cline{3-6}
 & & Data & Exp (signal) & Data & Exp (signal) \\ \hline
 & $e$ & $9$ & $4.32\pm 0.71$ $(82\%)$ & $3$ & $4.0\pm 0.6$ $(77\%)$ \\ \cline{2-6}
$e^+p$ & $\mu$ & $8$ & $3.70\pm 0.63$ $(92\%)$ & $3$ & $3.4\pm 0.5$ $(81\%)$ \\
\cline{2-6}
 & $\tau$ & $0$ & $0.82\pm 0.12$ $(46\%)$ & & \\ \hline
 & $e$ & $1$ & $3.18\pm 0.58$ $(70\%)$ & $3$ & $3.2\pm 0.5$ $(69\%)$ \\ \cline{2-6}
$e^-p$ & $\mu$ & $0$ & $2.40\pm 0.41$ $(92\%)$ & $2$ & $2.3\pm 0.4$ $(85\%)$ \\
\cline{2-6}
 & $\tau$ & $1$ & $0.68\pm 0.11$ $(31\%)$ & & \\ \hline
\end{tabular}
\label{tab:isol}
\end{center}
\end{table*}
H1 has observed an excess in both the $e$ and the $\mu$ channel. 
The excess is, however, not confirmed by ZEUS.
ZEUS has, on the other hand, observed 2 isolated $\tau$ lepton events with
a hadronic final state ($X$) at large transverse momentum 
($P^X_T>25\,{\rm GeV}$)
for $0.20\pm 0.05$ expected, based on the HERA-1 data, of which $49\%$
is contributed by the SM signal events from single $W$ production with
a genuine isolated $\tau$ and missing transverse momentum in the final state.
The corresponding signal contribution (purity) for the $e$ and $\mu$ channels 
is also shown in Table~\ref{tab:isol}.

\subsection{\boldmath Single $W$ production cross section}\label{wprod}

From the previous section, we noticed that the isolated lepton events with 
large $\not\!\! P_T$ are dominated by the SM $W$ production. 
It is possible, using the same sample, to further enhance the $W$ signal 
with additional selection cuts and to measure the $W$ 
production cross section.

The extracted production cross section from ZEUS~\cite{zeus08} is 
$\sigma(ep\rightarrow lWX)=0.89^{+0.25}_{-0.22} ({\rm stat})\pm 0.10 ({\rm syst})\,{\rm pb}$
at $\sqrt{s}=316\,{\rm GeV}$ corresponding to the luminosity weighted mean
of center of mass energies of the different data samples.
The corresponding theoretical expectation is $1.2\,{\rm pb}$ which
has an error of $15\%$ due to the uncertainty of the next to leading order 
corrections.

The corresponding measurement from H1 $\sigma (ep\rightarrow lWX)=1.14\pm 0.25({\rm stat)}\pm 0.14 ({\rm syst)}\,{\rm pb}$
is quoted for $\sqrt{s}=317\,{\rm GeV}$ and is in agreement with 
the prediction of $1.27\pm 0.19\,{\rm pb}$~\cite{h108}.

Both measurements from ZEUS and H1 have a significance of about five standard
deviations.

\subsection{\boldmath $W$ polarization fractions at HERA}

To test the compatibility of the observed $W$ decays with the SM, 
a measurement of the $W$ boson polarization is performed by H1~\cite{h108}.
The measurement is based on the $\cos \theta^\ast$ distribution 
in the decay $W\rightarrow e/\mu +\nu$, where $\theta^\ast$ is 
defined as the angle between the $W$ boson momentum in the lab frame and
that of the charged decay lepton in the $W$ boson rest frame.
The angular distribution is expected to have three distinct 
terms, $3/4F_0(-1-\cos^2\theta^\ast)$, $3/8F_-(1-\cos\theta^\ast)^2$
and $3/8F_+(1+\cos\theta^\ast)^2$ with $F_0$, $F_-$ and $F_+$ corresponding
to the longitudinal, left-handed and right-handed fractions respectively,
of which only two are independent as $F_0+F_-+F_+\equiv 1$.

\begin{figure}[htb]
\begin{center}
\includegraphics[scale=0.37]{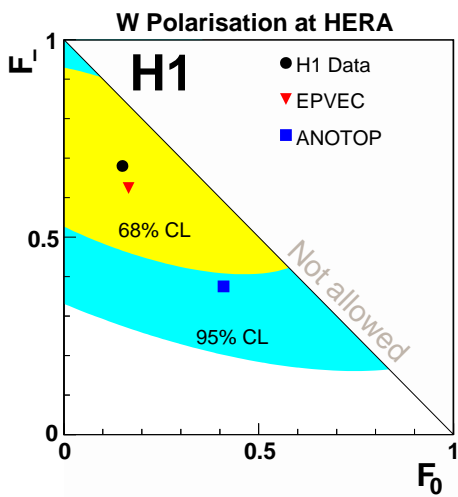}
\end{center}
\vspace{-10mm}
\caption{Preliminary $W$ boson polarization (point) at $1$ and $2\sigma$ CL 
(contours). Also shown are the values for the SM prediction (triangle) and
anomalous single top production via FCNC (square).}
\label{wfrac}
\end{figure}

The results on the polarization fractions $F_0$ and $F_-$ are obtained
from a simultaneous fit to the measured angular distribution and 
shown in Fig.~\ref{wfrac}.
The measurement is found in good agreement with the SM and compatible
with anomalous single top production via flavor changing neutral current
(FCNC, sec.~\ref{1top}).

\subsection{Single top production} \label{1top}

The production of single top quarks is kinematically possible at HERA.
The signature of a top decay to $b$ and $W$ with subsequent 
decay of the $W$ in the leptonic electron and muon channels
would be a lepton and
missing transverse momentum. 
This signature coincides with that of the observed
isolated lepton events discussed above.
The search is performed for anomalous top quark production
in an FCNC process involving the coupling $\kappa_{tu\gamma}$.
The dominant process for SM single top production at HERA has a cross 
section of less than $1\,{\rm fb}$ and is treated as background.

\begin{figure}[htb]
\begin{center}
\includegraphics[scale=0.37]{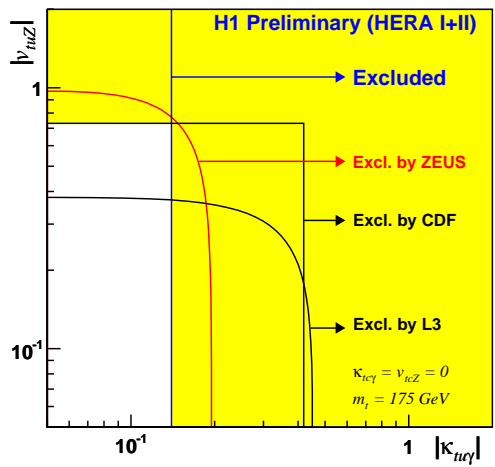}
\end{center}
\vspace{-10mm}
\caption{Exclusion limits at $95\%$ CL on the anomalous $\kappa_{tu\gamma}$
and $v_{tuZ}$ couplings obtained by the H1, ZEUS, CDF and L3 experiments.
The charm quark couplings $\kappa_{tc\gamma}$ and $v_{tcZ}$ are neglected, 
and the limits are shown assuming a top mass $m_t=175\,{\rm GeV}$.}
\label{fig:1top}
\end{figure}

The limits from HERA~\cite{h11top,zeus1top} are compared with 
other experiments~\cite{cdf1top,l31top} in Fig.~\ref{fig:1top}.
The H1 limit is based on the full HERA data, whereas the ZEUS limits 
are obtained from the HERA-1 data taking into account not only
$\kappa_{tu\gamma}$ but also $v_{tuZ}$ both in top production and decay.
ZEUS also studied the dependence of the limits on the top mass. It is
found that the limits are more stringent for a smaller top mass value, 
as expected.

\section{\boldmath GENERIC SEARCH AT HIGH $P_T$}\label{g-search}

\begin{figure}[htb]
\begin{center}
\includegraphics[angle=-90,scale=0.5]{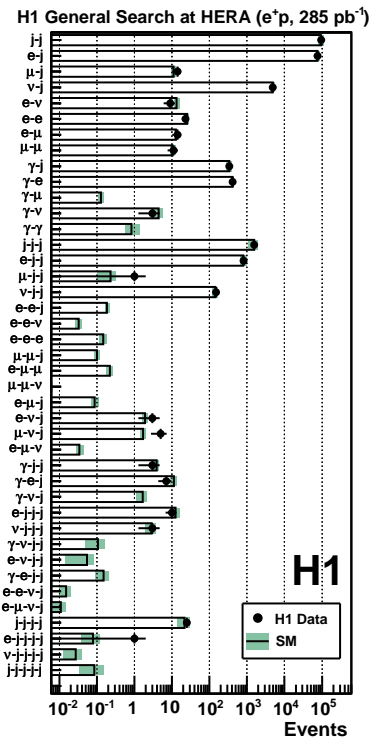}
\end{center}
\vspace{-10mm}
\caption{The data and the SM expectation for all event classes with
observed data events or a SM expectation greater than $0.01$ event.
The error bands on the predictions include model uncertainties and 
experimental systematic errors added in quadrature.}
\label{fig:generic}
\end{figure}

Different from specific searches presented so far, a generic and
model-independent search
for deviations from the SM prediction for all high transverse momentum
final state configuration involving electrons ($e$), muons ($\mu$),
jets ($j$), photons ($\gamma$) or neutrinos ($\nu$) was performed
by H1 first using the HERA-1 data~\cite{h1hera1} and 
now including the HERA-2 data~\cite{h1all}.
Events are classified in different classes. Each event class contains
at least two final state objects ($e,\mu,j,\gamma,\nu$) with
$P_T>20\,{\rm GeV}$
in the polar angle range $10^\circ <\theta <140^\circ$.
All event classes with observed data events or with a SM expectation greater
than $0.01$ event are shown for all H1 $e^+p$ data in Fig.~\ref{fig:generic}.
In a further step, a statistical algorithm is used to search for
deviations from the SM in the distributions of the scalar sum of transverse
momenta or invariant mass of final state particles.
%The largest deviation is observed in the $\mu-j-\nu$ event class
%in $e^+p$ data coinciding with the observation in Sec.~\ref{sec:isol}.

\section{SUMMARY}

The HERA $ep$ collider is first of all a precision QCD machine providing 
dominant constraints for parton distribution functions at low $x$.
It is also a high energy collider complementary to LEP $e^+e^-$ and 
the Tevatron $p\bar{p}$ colliders allowing rich electroweak 
measurements as well as many searches for new particles to be performed.
 
There is a good prospect to achieve much improved electroweak results 
in the next two years or so before the next Ringberg workshop. 
The $W$ propagator measurement will be mainly improved using
the ten-fold increases in the $e^-p$ data sample at HERA-2.
The light quark couplings to the $Z$ boson will be further constrained
using the complete polarized HERA-2 data.

Most searches are now based on the full
HERA data although some of them are still in preliminary form.
In essentially all cases, no significant deviation with the SM expectation
is observed, with one possible exception, namely the excess observed
by H1 on the isolated electron and muon events with large missing transverse 
energy, which has around $2.4$ standard deviations. 
This excess is, however, not confirmed by a similar analysis from ZEUS. 
The various limits derived from the HERA data are often comparable to 
or more stringent than those from LEP and the Tevatron experiments.

\end{document}